\documentclass[prb,aps,twocolumn,superscriptaddress,floatfix,citeautoscript]{revtex4-1}
\usepackage{multirow}         
\usepackage{threeparttable}   
\usepackage{dcolumn}          
\usepackage[latin1]{inputenc} 
\usepackage[dvips]{graphicx}

\def\x{$\times$}

\begin{document}
\title{A Higher-Accuracy van der Waals Density Functional}
\author{Kyuho Lee}
\affiliation{Department of Physics and Astronomy, Rutgers University,
  Piscataway, New Jersey 08854-8019, USA}
\author{\'{E}amonn D. Murray}
\affiliation{Department of Physics and Astronomy, Rutgers University,
  Piscataway, New Jersey 08854-8019, USA}
\author{Lingzhu Kong}
\affiliation{Department of Physics and Astronomy, Rutgers University,
  Piscataway, New Jersey 08854-8019, USA}
\author{Bengt I. Lundqvist}
\affiliation{Department of Applied Physics, Chalmers University of
  Technology, SE - 41296 G\"{o}teborg, Sweden}
\affiliation{Center for Atomic-scale Materials Design, Department of
  Physics \\ Technical University of Denmark, DK - 2800 Kongens
  Lyngby, Denmark}
\author{David C. Langreth}
\affiliation{Department of Physics and Astronomy, Rutgers University,
  Piscataway, New Jersey 08854-8019, USA}
\date{\today}
\begin{abstract}
  We propose a second version of the van der Waals density functional
  (vdW-DF2) of Dion et al. [Phys.\ Rev.\ Lett.\ \textbf{92}, 246401
  (2004)], employing a more accurate semilocal exchange functional and
  the use of a large-$N$ asymptote gradient correction in determining
  the vdW kernel.  The predicted binding energy, equilibrium
  separation, and potential-energy curve shape are close to those of
  accurate quantum chemical calculations on 22 duplexes.  We
  anticipate the enabling of chemically accurate calculations in
  sparse materials of importance for condensed-matter, surface,
  chemical, and biological physics.
\end{abstract}
\maketitle

The van der Waals (vdW) attraction is a quantum-mechanical phenomenon
with charge fluctuations in one part of an atomic system that are
electrodynamically correlated with charge fluctuations in another. The
vdW force at one point thus depends on charge events at another region
and is a truly nonlocal correlation effect.

Methods for accurately calculating the vdW interactions are critical
to understanding sparse matter, including bulk solids ({\it e.g.},
layered materials, molecular crystals, and polymers), surface
phenomena ({\it e.g.}, adsorption, water overlayers, and gas
separation and storage), and biostructures ({\it e.g.}, DNA and
protein structure).

The exact density functional contains the vdW forces. Unfortunately,
we do not have access to it, but approximate versions are
abundant. Commonly, the local-density approximation (LDA) and
generalized gradient approximation (GGA) are used with quite some
success for dense matter, including hard materials and covalently
bound molecules. They depend on the density in local and semilocal
ways, respectively, however, and give no account of the fully nonlocal
vdW interaction.

First-principles approaches for how vdW can be treated in DFT were
first proposed for the asymptotic interaction between
fragments~\cite{Andersson1996,Dobson1996,Kohn1998b}.  These ultimately
evolved into the van der Waals density functional (vdW-DF) for
arbitrary geometries~\cite{Dion2004,Thonhauser2007,Vydrov}.  Despite
its success for describing dispersion in a breadth of systems better
than any other nonempirical method~\cite{Langreth2009}, vdW-DF
overestimates equilibrium separations ~\cite{Dion2004, Puzder2006,
  Chakarova-Kack2006, Thonhauser2007, Kong2009, Langreth2009,
  Toyoda2009a, Romaner2009} and underestimates hydrogen-bond
strength~\cite{Gulans2009,Kelkkanen2009}.

In this Letter, we propose a second version of the van der Waals
density functional (vdW-DF2) employing a more accurate semilocal
exchange functional PW86~\cite{PW86,PW86R} and the use of a large-$N$
asymptote gradient correction~\cite{Elliott2009} in determining the
vdW kernel.  By making a full comparison of potential energy curves
(PECs) with accurate quantum chemistry results for 22 molecular
duplexes, we show that vdW-DF2 substantially improves (i) equilibrium
separations, (ii) hydrogen bond strengths, and (iii) vdW attractions
at intermediate separations longer than the equilibrium ones.  The
improvement in (iii), found via a full PEC comparison, is most
critical for important ``real-life'' applications to sparse matter and
biological matter where it is impossible for basic structural units to
assume the same separations they would have as binary units in vacuo.

First, we replace revPBE exchange functional~\cite{revPBE} with
PW86~\cite{PW86,PW86R}, because revPBE is generally too repulsive near
the equilibrium separation~\cite{Puzder2006}, and can bind spuriously
by exchange alone, although less so than most other local or semilocal
functionals. Hence, other exchange functionals
~\cite{Cooper2010,Klimes2010} have been proposed.  Recent performance
studies of various exchange functionals for weakly interacting
atoms~\cite{KB09} and molecules~\cite{Murray2009}, however, show PW86,
with an enhancement factor proportional to $s^{2/5}$ at large reduced
density gradient $s$, to give the most consistent agreement with
Hartree-Fock (HF) results, without spurious exchange binding. It also
is a good match~\cite{vdW-DF-PW86} for the vdW-DF2 correlation kernel,
introduced below, although no others were tried.

The key to the vdW-DF method is the inclusion of a long range piece of
the correlation energy, $E^{\mathrm{nl}}_c[n]$, a fully nonlocal
functional of the density $n$.  This piece is evaluated using a
``plasmon'' pole approximation for the inverse dielectric function,
which satisfies known conservation laws, limits, sum rules, and
invariances~\cite{Dion2004}. A single parameter model for the pole
position was adopted, with the pole residue set by the law of
charge-current continuity ($f$-sum rule), and the pole position at
large wave vector set by the constraint that there be no self-Coulomb
interaction. The single parameter is determined locally from
electron-gas energy input via gradient corrected LDA~\cite{Dion2004}.

\begin{figure*}[!t]
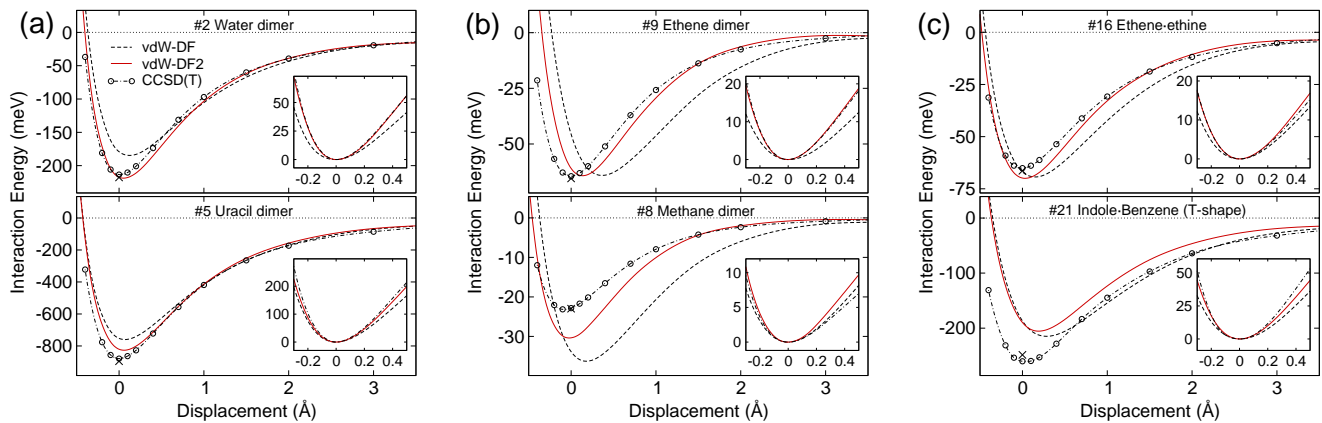

  \centering
  \includegraphics[height=2.1666in]{fig1a.eps}\hspace{5mm}
  \includegraphics[height=2.1666in]{fig1b.eps}\hspace{5mm}
  \includegraphics[height=2.1666in]{fig1c.eps}
  \caption{(Color online) Potential energy curves (PECs) for the best
    and the worst case of (a) hydrogen-bonded, (b)
    dispersion-dominated, and (c) mixed duplexes. CCSD(T) QC PECs
    (dash-dotted lines with circles taken from
    Ref.~\onlinecite{Molnar2009}) and the reference energies (cross
    marks taken from Ref.~\onlinecite{Takatani2010}) at the geometry
    of Ref.~\onlinecite{Jurecka2006} are also shown. The shapes near
    minima are compared in inset figures where PECs are aligned to
    have the common minimum point. For all the other S22 duplexes, see
    Supplemental Material.}
  \label{fig:pecs}
\end{figure*}

The nonlocal piece of the correlation energy in both vdW-DF and
vdW-DF2 is of the form
\newcommand{\rr}{\mathbf{r}}
\begin{equation}
\label{phiint}
 E_c^{nl}[n]=\int\! d^3r\!\int\! d^3r'\,n(\rr)
\phi(\rr,\rr')n(\rr').
\end{equation}
The kernel $\phi$ is given as a function of $Rf(\rr)$ and $Rf(\rr')$,
where R=$|\rr-\rr'|$ and $f(\rr)$ is a function of $n(\rr)$ and its
gradient. In fact $f(\rr)$ is proportional to the exchange-correlation
energy density $\epsilon_{xc}$ of a gradient corrected LDA at the
point $\rr$. This arose from the approximately implemented requirement
that the dielectric function implied by the plasmon pole model should
give an exchange-correlation energy semilocally consistent with a
gradient corrected electron gas.  We call the semilocal functional
that fixes $f(\rr)$ in Eq.~(\ref{phiint}) the \textit{internal
  functional}.

The above is easier to understand for two separate molecules, although
the arguments apply equally well to a pair of high density regions of
a sparse material. The long range vdW attraction implied by
Eq.~(\ref{phiint}) occurs from the contribution where $\rr$ is on one
molecule and $\rr'$ on the other.  The definition of $f(\rr)$ and
$f(\rr')$ varies continuously and independently at each point
according to $\epsilon_{xc}(\rr)$ and $\epsilon_{xc}(\rr')$. The
quantity $\epsilon_{xc}$ is taken to consist of a gradient corrected
LDA.  In the first version of vdW-DF~\cite{Dion2004}, the gradient
correction was obtained from a gradient expansion~\cite{LV1990} for
the slowly varying electron gas~\cite{Zdiscussion,Zdef}.  More
appropriate is a functional that gives accurate energies for
\textit{molecules}, however.  When $\rr$ and $\rr'$ are each in a
separate molecule-like region, with exponentially decaying tails
between them, $f(\rr)$ and $f(\rr')$ can both be large and give key
contributions to a vdW attraction.  For this case (including perhaps
even a molecule near a surface) the large-$N$
asymptote~\cite{Schwinger1980,Schwinger1981} and the exchange energy
asymptotic series for neutral atoms provide a more accurate
approximation.  In fact, the exchange parameter~\cite{Zdef} $\beta$ of
the B88 exchange functional~\cite{B88}, successfully used in the vast
majority of DFT calculations on molecules, can be derived from first
principles using the large-$N$ asymptote~\cite{Elliott2009}, as can
the LDA exchange.  It seems obvious, then, that vdW-DF results should
be improved if the second order expansion of the exchange in gradients
is replaced by the second order large-$N$ expansion. Interestingly,
PW86R functional, selected as the overall exchange functional for
different reasons, also follows the large-$N$ behavior for small
reduced gradient $s$ values down to $\sim$0.1, where it reverts to the
form of slowly varying electron gas limit.

Thus we use 2.222 times larger exchange gradient coefficient, a value
based on agreement between
derived~\cite{Schwinger1980,Schwinger1981,Elliott2009} and
empirical~\cite{B88} criteria (a 6\% smaller derived value of
Ref.~\onlinecite{Elliott2009} only gives a marginal improvement).
Assuming that the \textit{screened} exchange term~\cite{Screening}
increases in the same proportion as gradient exchange itself, finally
we get the appropriate gradient coefficient in the ``Z''
notation~\cite{Zdef} which is multiplied by 2.222.  Summarizing: while
$Z_{ab}=-0.8491$ in vdW-DF, $Z_{ab}=-1.887$ in vdW-DF2, implying
changes in the internal functional.

The performance of our new energy functional is assessed via
comparisons with the accurate S22 reference
dataset~\cite{Jurecka2006,Takatani2010} and PECs~\cite{Molnar2009}
based on quantum chemistry (QC) calculations at the level of CCSD(T)
with extrapolation to the complete basis set limit. These twenty-two
small molecular duplexes for the non-covalent interactions typical in
biological molecules include hydrogen-bonded, dispersion-dominated,
and mixed duplexes.  Recent evaluation~\cite{Gulans2009} of the
performance of the vdW-DF for S22 shows it to be quite good, except
for H-bonded duplexes, where vdW-DF underestimates the binding energy
by about 15\%.

Calculations are performed by a plane-wave code and an efficient vdW
algorithm~\cite{Soler} with Troullier-Martins type norm-conserving
pseudopotentials.  Spot comparison with all electron calculations
using large basis sets indicates a calculational accuracy of
$\sim$1\%, actually better than that of most of PAW potentials
supplied in various standard codes.  Large box sizes were used to
control spurious electrostatic interactions between replicas.  See
Supplemental Material for further details.

\begin{figure}[!b]
  \centering
  \includegraphics[width=3in,clip=true]{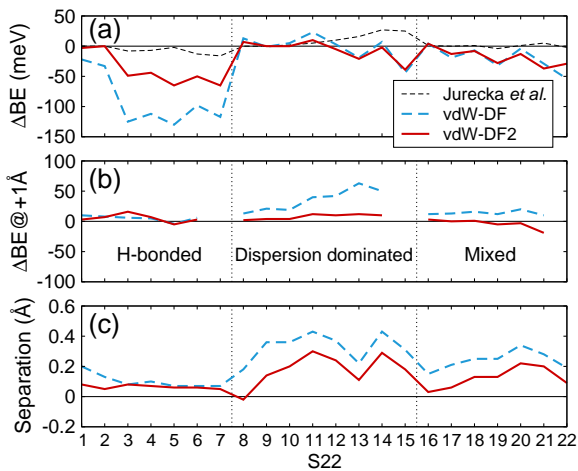}
  \caption{(Color online) Comparison for the S22 duplexes of (a)
    binding energy predicted by Jurecka et al.~\cite{Jurecka2006},
    vdW-DF~\cite{Dion2004}, and the present work (vdW-DF2).  (b)
    binding energy at a separation 1 \AA{} larger than the equilibrium
    one and, (c) equilibrium separations.  The ordinates give the
    respective deviations of these quantities from the reference
    values taken from Takatani et al.~\cite{Takatani2010} for panel
    (a), Molnar et al.~\cite{Molnar2009} for panel (b), and Jurecka et
    al.~\cite{Jurecka2006} for panel (c).}
  \label{fig:comp} 
\end{figure}

Figure~\ref{fig:pecs} shows a typical PEC for each kind of
interaction.  To make a direct comparison to the QC results, the PECs
are calculated at the same geometry of those of the CCSD(T) PEC
calculations~\cite{Molnar2009} (shown as dash-dotted lines with
circles), where each molecule is kept in its S22
geometry~\cite{Jurecka2006} and moved along the line defined by the
center-of-mass coordinates of two molecules without any rotation.
Overall, the vdW-DF2 PECs (solid lines) show a remarkable agreement
with QC ones for all separations and for all three types of
interactions.  The shapes of the PECs near the minima, important for
vibrational frequencies, are greatly improved (see inset figures where
PECs are aligned to have the common minimum point).  More importantly
the strength of vdW attraction at larger distances is weakened in
agreement with QC, especially for vdW-bonded duplexes
(Fig.~\ref{fig:pecs}(b)).  In other words, the original vdW-DF
overestimates vdW attraction at intermediate separations beyond the
equilibrium separation, although its minimum energy is accurate.  This
has special importance in sparse condensed matter. An example will be
given below.

The PECs tend to turn up slightly earlier when approaching the
repulsion regime at small separations.  This quite universal feature
might be due to PW86 exchange functional which is slightly more
repulsive than HF at short distances~\cite{Murray2009}.  For duplexes
whose large distance asymptote is dominated by dispersion (here
methane dimer, ethene dimer, and benzene--methane), vdW-DF2 will have
weaker attraction in the asymptotic region and smaller $C_6$
coefficients than vdW-DF, which (at least for the methane dimer)
already gives a $C_6$ coefficient close to experiment~\cite{ Vydrov}.
Recently, this prediction was verified in detail for various
duplexes~\cite{VV10}.  However, in the region of our calculations,
such deterioration is not pervasive.  In any case, neither the vdW-DF
curves nor the QC curve have reached their asymptotic limit for any of
the above three cases.  The remainder of the 22 duplexes have
asymptotic forms dominated by electrostatics.

In Fig.~\ref{fig:comp}, we summarize the calculated binding energies
at equilibrium separation, at equilibrium separation plus 1 {\AA}, and
the equilibrium separations themselves, with each quantity given as a
deviation from that implied by the reference
calculations~\cite{Jurecka2006,Molnar2009,Takatani2010}.  Each
subfigure clearly shows one of the three major improvements:
\textit{(i) Hydrogen-bond strengths}. The mean absolute deviation
(MAD) of binding energy for hydrogen-bonded duplexes is decreased from
91 to 40 meV.  \textit{(ii) Interaction at intermediate separations.}
The MAD of binding energy at a separation 1 \AA{} larger than the
equilibrium one is reduced from 35 to 8 meV for the dispersion
dominated group and also substantially for the mixed group.
\textit{(iii) Equilibrium separations.} MAD is reduced from 0.23 to
0.13 {\AA}.  \textit{Overall}, the vdW-DF2 binding energies are within
50 meV of the reference except for duplexes 5 and 7. The MAD of
binding energy is decreased from 41 to 22 meV (13\% to 7.6\%). As a
final tidbit, we note that in vdW-DF2 the MAD for the equilibrium
energy of the dispersion dominated complexes has been reduced to 11
meV, which is equal to the MAD of the work of Jurecka et
al.~\cite{Jurecka2006}, which until very recently~\cite{Takatani2010}
was considered the `gold standard' for accuracy in quantum chemical
calculations on this group.

The higher accuracy of vdW-DF2 persists in the extended systems we
tested. As such applications, we calculated: (i) The graphite
interlayer binding energy and spacing. The binding energy is on a par
with that of vdW-DF which is already good. Interlayer spacing is about
2\% shorter than that by vdW-DF, in better agreement with experiment.
(ii) H$_2$ adsorption within two different metal-organic frameworks
(MOFs), Zn$_2$(BDC)$_2$(TED) and MOF-74. In the former, the binding at
the highest binding site (57 meV including zero point) is in good
agreement with heat of adsorption measurement (52 meV), whereas vdW-DF
overestimates it by 60\%.  This demonstrates the importance of
accurate intermediate-range interaction. For comparison, the binding
energy of hydrogen to benzene, one of building blocks of
Zn$_2$(BDC)$_2$(TED), is almost the same for both functionals at the
equilibrium separation.  At larger separations, however, vdW-DF
substantially overestimates the binding.  In case of MOF-74, H$_2$
positions are experimentally known.  VdW-DF2 shows its higher accuracy
in predicting separations between hydrogens and nearby atoms (improved
by 0.1-0.2 \AA{}). The strongest binding comes from an unsaturated
metal atom, rather than more distant structures, and the binding
energy, already accurate in vdW-DF, is little changed in vdW-DF2.  We
also tested vdW-DF2 on the internal structure of a water molecule, as
an example of strong chemical bonds in molecules. We find that vdW-DF2
is on par with PBE.  For details of all tests, see Supplemental
Material.

In summary, we have presented an enhanced version of vdW-DF, denoted
by vdW-DF2, which can be implemented via simple modifications of
existing codes. It results in significant improvements in equilibrium
spacings between noncovalently bound complexes, as well as in binding
energy, especially when hydrogen bonding plays a role.  We make a full
comparison of PECs of both functionals with accurately known results
for a set of 22 complexes and also apply it to extended solid systems,
finding favorable results for the new functional. Thus, we expect our
method to have important applications in a wide range of fields,
including condensed matter and materials physics, chemical physics,
and the physics of biological materials.

We thank V. Cooper, A. Gulans, and R. Nieminen for discussions,
P. Elliott and K. Burke for an early copy of
Ref.~\onlinecite{Elliott2009}, and D. Case for the use of his computer
cluster for several spot checks.  Work principally supported by
NSF-DMR-0801343; work on MOFs by DE-FG02-08ER46491; DCL work at KITP
by NSF-PHY05-51164; BIL work by the Lundbeck Foundation via CAMD.


\setcounter{figure}{0}
\begin{center}
\textbf{Supplemental Material}
\end{center}
\textit{Computational details} The convergence control parameters
for the plane-wave pseudopotential calculations are tuned until the
binding energy is converged up to 0.4 meV (0.01 kcal~mol$^{-1}$). We used 50
Rydberg kinetic-energy cutoff for hydrocarbon systems and 60 Rydberg
for all the others containing nitrogen and oxygen. For polar molecules
with large dipole moments a 60\x60\x60 Bohr$^3$ cubic unit cell is
needed to eliminate spurious electrostatic interaction between
supercell images.  The accuracy of the pseudopotentials are tested
within PBE [J. Perdew, K. Burke, and M. Ernzerhof,
Phys. Rev. Lett. {\bf 77}, 3865 (1996)] by comparing with all-electron
results for the water dimer (duplex \#2 in S22) and formic acid dimer
(duplex \#3).  The all-electron energies obtained were also converged
to the level of 0.4 meV (See Table~\ref{tab:g03}).  For water dimer we
obtained binding energies (in meV) of 215 (all electron) and 219
(pseudopotential).  Similarly, for the formic acid dimer we obtained
790 (all electron) and 798 (pseudopotential). The respective
deviations are thus 2\% and 1\%.
 
Compared to GGA, the vdW-DF2 computation time per iteration roughly
doubled for the smaller S22 duplex calculations, but the increase
becomes virtually unmeasurable for the larger MOF systems that have
more than 100 atoms per unit cell.  The number of iterations required
for convergence is also typically comparable to that of GGA, but in
some MOF cases it is almost doubled.

\begin{table}[!h]
  \centering
  \caption{Convergence of all-electron PBE binding energy (in meV)
    of the water dimer (duplex \#2 in S22) and the formic acid dimer
    (duplex \#3).  The calculations are performed by Gaussian03.$^a$
    Dunning's correlation consistent basis sets with diffuse functions
    are used, with counterpoise-corrections.}
  \label{tab:g03}
  \begin{ruledtabular}
    \begin{tabular}{ldd}
 basis set & 
\multicolumn{1}{c}{water dimer} & 
\multicolumn{1}{c}{formic acid dimer} \\ \hline
aug-cc-pVDZ & 212 & 774 \\
aug-cc-pVTZ & 213 & 784 \\
aug-cc-pVQZ & 214 & 790 \\
aug-cc-pV5Z & 215 & 790 \\
aug-cc-pV6Z & 215 & \multicolumn{1}{c}{--} \\
    \end{tabular}
  \end{ruledtabular}
  \footnotetext[1]{Gaussian 03, Revision E.01,
 M. J. Frisch, G. W. Trucks, H. B. Schlegel, G. E. Scuseria,
 M. A. Robb, J. R. Cheeseman, J. A. Montgomery, Jr., T. Vreven,
 K. N. Kudin, J. C. Burant, J. M. Millam, S. S. Iyengar, J. Tomasi,
 V. Barone, B. Mennucci, M. Cossi, G. Scalmani, N. Rega,
 G. A. Petersson, H. Nakatsuji, M. Hada, M. Ehara, K. Toyota,
 R. Fukuda, J. Hasegawa, M. Ishida, T. Nakajima, Y. Honda, O. Kitao,
 H. Nakai, M. Klene, X. Li, J. E. Knox, H. P. Hratchian, J. B. Cross,
 V. Bakken, C. Adamo, J. Jaramillo, R. Gomperts, R. E. Stratmann,
 O. Yazyev, A. J. Austin, R. Cammi, C. Pomelli, J. W. Ochterski,
 P. Y. Ayala, K. Morokuma, G. A. Voth, P. Salvador, J. J. Dannenberg,
 V. G. Zakrzewski, S. Dapprich, A. D. Daniels, M. C. Strain,
 O. Farkas, D. K. Malick, A. D. Rabuck, K. Raghavachari,
 J. B. Foresman, J. V. Ortiz, Q. Cui, A. G. Baboul, S. Clifford,
 J. Cioslowski, B. B. Stefanov, G. Liu, A. Liashenko, P. Piskorz,
 I. Komaromi, R. L. Martin, D. J. Fox, T. Keith, M. A. Al-Laham,
 C. Y. Peng, A. Nanayakkara, M. Challacombe, P. M. W. Gill,
 B. Johnson, W. Chen, M. W. Wong, C. Gonzalez, and J. A. Pople,
 Gaussian, Inc., Wallingford CT, 2004.}
\end{table}

\begin{table}[!h]
  \centering
  \caption{Internal structure of a free water molecule.  The bond
    length and bond angle calculated by vdW-DF2, vdW-DF, and PBE
    well agree each other within 0.002 \AA{} and 0.7 degree.
    Norm-conserving pseudopotentials (NCPP), 20\x20\x20 Bohr$^3$ cubic
    unit cell, 130 Rydberg kinetic-energy cutoff, and 0.002 eV/\AA\
    force tolerance are used. For comparison all-electron (AE) PBE,
    CCSD(T), and experiment values are given as well. AE PBE bond
    length is 0.007--0.009 \AA{} longer than our pseudopotential
    calculation.}
  \label{tab:h2o}
  \begin{ruledtabular}
    \begin{tabular}{cdddddd}
      Method                  & \multicolumn{1}{c}{$d$(OH) (\AA)} &
      \multicolumn{1}{c}{$\theta$(HOH) ($^\circ$)} \\
      \hline
      vdW-DF2 (NCPP)          & 0.960   & 105.0 \\
      vdW-DF  (NCPP)          & 0.960   & 104.7 \\
      PBE     (NCPP)          & 0.962   & 104.3 \\
      PBE (AE)                & 0.969\footnotemark[1],
                                0.971\footnotemark[2]
                              & 104.1\footnotemark[2] \\
      CCSD(T)\footnotemark[3] & 0.958   & 104.5 \\
      Expt.\footnotemark[4]   & 0.958   & 104.5 \\
    \end{tabular}
  \end{ruledtabular}
  \footnotetext[1]{
    All-electron calculation with the def2-TZVPP basis set
    (triple-$\zeta$ quality including high exponent polarization
    functions), which is a larger basis set than the cc-pCVTZ
    basis set, taken from J. Chem. Phys. {\bf 126}, 124115 (2007). 
  }
  \footnotetext[2]{
    All-electron calculation with aug-cc-pVTZ basis set taken from
    J. Phys. Chem. A {\bf 108}, 2305 (2004).
  }
  \footnotetext[3]{
    D.~Feller and K.~A.~Peterson, J. Chem. Phys. {\bf 131}, 154306 (2009).
  }
  \footnotetext[4]{
    S.~V.~Shirin {\it et al.}, J. Mol. Spectrosc. {\bf 236}, 216 (2006).
  }
\end{table}

\clearpage
\begin{table*}[!ht]
\begin{minipage}{\textwidth}
  \centering
  \caption{Comparison of binding energy and equilibrium
    separation. The latter is given as a deviation from the original
    S22 geometry$^a$ along the center-of-mass line, as shown in the
    first data column of Table~\ref{tab:cm-s22}).}
  \label{tab:comp-s22}
  \begin{ruledtabular}
    \begin{tabular}{ccdddddd}
\multirow{2}{*}{\#} & 
\multirow{2}{*}{Duplex} & 
\multicolumn{4}{c}{Binding energy (meV)} & 
\multicolumn{2}{c}{Separation (\AA)} \\

      \cline{3-6} \cline{7-8}

   &        & \multicolumn{1}{c}{vdW-DF} & \multicolumn{1}{c}{vdW-DF2}
   & \multicolumn{1}{c}{QC\footnotemark[1]}
   & \multicolumn{1}{c}{QC\footnotemark[2]}
   & \multicolumn{1}{c}{vdW-DF} & \multicolumn{1}{c}{vdW-DF2} \\

\hline

 1 & Ammonia dimer                      & 115 & 134 & 137 & 137 & 0.20 &  0.08 \\
 2 & Water dimer                        & 185 & 218 & 218 & 218 & 0.13 &  0.05 \\
 3 & Formic acid dimer                  & 690 & 766 & 807 & 815 & 0.08 &  0.08 \\
 4 & Formamide dimer                    & 587 & 655 & 692 & 699 & 0.10 &  0.07 \\
 5 & Uracil dimer                       & 767 & 832 & 895 & 897 & 0.07 &  0.06 \\
 6 & 2-pyridoxine$\cdot$2-aminopyridine & 639 & 687 & 725 & 737 & 0.07 &  0.06 \\
 7 & Adenine$\cdot$thymine              & 609 & 660 & 710 & 726 & 0.07 &  0.05 \\
\hline
 8 & Methane dimer                      &  36 &  30 &  23 &  23 & 0.18 & -0.02 \\
 9 & Ethene dimer                       &  64 &  65 &  65 &  65 & 0.36 &  0.14 \\
10 & Benzene$\cdot$methane              &  68 &  63 &  65 &  63 & 0.36 &  0.20 \\
11 & Benzene dimer (slip-parallel)      & 136 & 123 & 118 & 114 & 0.43 &  0.30 \\
12 & Pyrazine dimer                     & 185 & 177 & 192 & 182 & 0.36 &  0.24 \\
13 & Uracil dimer (stacked)             & 403 & 402 & 439 & 422 & 0.22 &  0.11 \\
14 & Indole$\cdot$benzene (stacked)     & 206 & 197 & 226 & 199 & 0.42 &  0.29 \\
15 & Adenine$\cdot$thymine (stacked)    & 461 & 466 & 530 & 506 & 0.30 &  0.18 \\
\hline
16 & Ethene$\cdot$ethine                &  69 &  70 &  66 &  65 & 0.15 &  0.03 \\
17 & Benzene$\cdot$water                & 124 & 129 & 142 & 143 & 0.20 &  0.06 \\
18 & Benzene$\cdot$ammonia              &  94 &  92 & 102 & 101 & 0.27 &  0.13 \\
19 & Benzene$\cdot$HCN                  & 166 & 170 & 193 & 197 & 0.24 &  0.13 \\
20 & Benzene dimer (T-shape)            & 113 & 105 & 119 & 118 & 0.34 &  0.22 \\
21 & Indole$\cdot$benzene (T-shape)     & 214 & 206 & 248 & 243 & 0.28 &  0.20 \\
22 & Phenol dimer                       & 254 & 279 & 306 & 307 & 0.19 &  0.09 \\
\hline
   & Mean deviation (MD)                & -36 & -21 &   2 &     & 0.23 &  0.13 \\
\hline
   & Mean absolute deviation (MAD)      &  41 &  22 &   7 &     & 0.23 &  0.13 \\
   & MAD\%
   & \multicolumn{1}{c}{13\%}
   & \multicolumn{1}{c}{8\%}
   & \multicolumn{1}{c}{2\%} &  &      &       \\
    \end{tabular}
  \end{ruledtabular}
  \footnotetext[1]{P.~Jurecka, J.~Sponer, J.~Cern\'{y}, and P.~Hobza,
    Phys. Chem. Chem. Phys. {\bf 8}, 1985 (2006).}
  \footnotetext[2]{T.~Takatani, E.~G.~Hohenstein, M. Malagoli,
    M.~S.~Marshall, and C.~D.~Sherrill, J. Chem. Phys. {\bf 132},
    144104 (2010).}
\end{minipage}
\end{table*}

\begin{table*}[!ht]
\begin{minipage}{\textwidth}
  \centering
  \caption{Left two data columns: The center-of-mass separations and
    the nearest neighbor (NN) pair separations in the S22 geometry of
    Jurecka \textit{et al}. These center-of-mass distances corresponds
    to the zeros of the abscissa in the PEC plots in
    Figs.~\ref{fig:h-pecs}--\ref{fig:comp-pw86}. The species of NN
    atom pair are given in the parentheses.  Right two data columns:
    The deviations from the CCSD(T) value [J. Chem.\ Phys.~{\bf 131},
    065102 (2009)] of the binding energy at a distance 1 \AA\ larger
    than that of the S22 equilibrium geometry of Jurecka \textit{et
      al}. [Phys. Chem. Chem. Phys. {\bf 8}, 1985 (2006)].}
  \label{tab:cm-s22}
  \begin{ruledtabular}
    \begin{tabular}{ccccdd}
      \multirow{2}{*}{\#} & 
      \multirow{2}{*}{Duplex} & 
      \multicolumn{2}{c}{Separation (\AA)} &
      \multicolumn{2}{c}{$\Delta$BE at +1 \AA{} (meV)} \\
      \cline{3-4}      \cline{5-6}
      &        & 
      \multicolumn{1}{c}{Center-of-mass} &
      \multicolumn{1}{c}{NN pair} &
      \multicolumn{1}{c}{vdW-DF} &
      \multicolumn{1}{c}{vdW-DF2} \\
      \hline                                  
       1 & Ammonia dimer                      & 3.21 & 2.50 \text{ (NH)} & 10 &  3 \\
       2 & Water dimer                        & 2.91 & 1.95 \text{ (OH)} &  8 &  7 \\
       3 & Formic acid dimer                  & 2.99 & 1.67 \text{ (OH)} &  6 & 16 \\
       4 & Formamide dimer                    & 3.23 & 1.84 \text{ (OH)} &  5 &  7 \\
       5 & Uracil dimer                       & 6.07 & 1.77 \text{ (OH)} & -4 & -5 \\
       6 & 2-pyridoxine$\cdot$2-aminopyridine & 5.14 & 1.86 \text{ (NH)} &  6 &  3 \\
       7 & Adenine$\cdot$thymine              & 5.97 & 1.82 \text{ (NH)} &
       \multicolumn{1}{c}{~N/A\footnotemark[1]}& \multicolumn{1}{c}{~N/A\footnotemark[1]}\\
       \hline
       8 & Methane dimer                      & 3.72 & 3.51 \text{ (CH)} & 13 &  2 \\
       9 & Ethene dimer                       & 3.72 & 2.56 \text{ (HH)} & 21 &  4 \\
      10 & Benzene$\cdot$methane              & 3.72 & 2.79 \text{ (CH)} & 19 &  4 \\
      11 & Benzene dimer (slip-parallel)      & 3.76 & 3.37 \text{ (CC)} & 40 & 12 \\
      12 & Pyrazine dimer                     & 3.48 & 3.27 \text{ (NH)} & 42 & 10 \\
      13 & Uracil dimer (stacked)             & 3.17 & 2.71 \text{ (OH)} & 63 & 12 \\
      14 & Indole$\cdot$benzene (stacked)     & 3.50 & 3.20 \text{ (CH)} & 50 & 10 \\
      15 & Adenine$\cdot$thymine (stacked)    & 3.17 & 2.68 \text{ (HH)} &
      \multicolumn{1}{c}{~N/A\footnotemark[1]}&\multicolumn{1}{c}{~N/A\footnotemark[1]}\\
      \hline
      16 & Ethene$\cdot$ethine                & 4.42 & 2.83 \text{ (CH)} & 12 &   3 \\
      17 & Benzene$\cdot$water                & 3.38 & 2.60 \text{ (CH)} & 13 &   0 \\
      18 & Benzene$\cdot$ammonia              & 3.56 & 2.77 \text{ (CH)} & 16 &   1 \\
      19 & Benzene$\cdot$HCN                  & 3.95 & 2.67 \text{ (CH)} & 12 &  -5 \\
      20 & Benzene dimer (T-shape)            & 4.91 & 2.80 \text{ (CH)} & 20 &  -3 \\
      21 & Indole$\cdot$benzene (T-shape)     & 4.88 & 2.59 \text{ (CH)} & 10 & -19 \\
      22 & Phenol dimer                       & 4.92 & 1.94 \text{ (OH)} &
      \multicolumn{1}{c}{~N/A\footnotemark[1]}& \multicolumn{1}{c}{~N/A\footnotemark[1]}\\
      \hline
      & Mean deviation (MD)                &       &       & 19 & 3 \\ \hline
      & Mean absolute deviation (MAD)      &       &       & 19 & 7 \\
      & MAD\%                              &       &
      & \multicolumn{1}{c}{16\%} & \multicolumn{1}{c}{4\%} \\
    \end{tabular}
  \end{ruledtabular}
  \footnotetext[1]{
    The CCSD(T) PEC is not available.
  }
\end{minipage}
\end{table*}

\begin{figure*}[!ht]
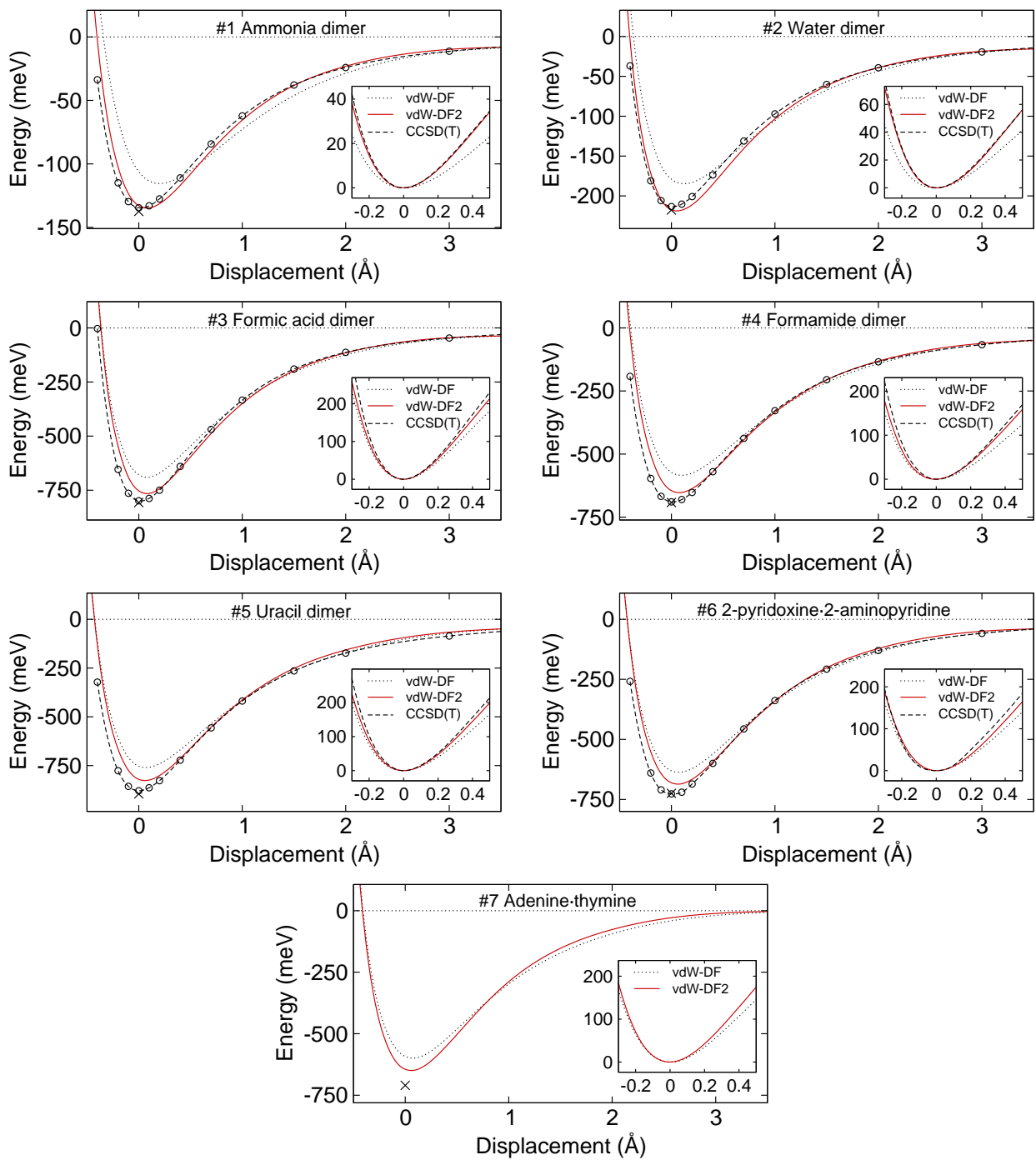

  \centering
  \includegraphics*[width=0.44\textwidth,clip]{01.eps}\hspace{5mm}
  \includegraphics*[width=0.44\textwidth,clip]{02.eps}\\\vspace{2mm}
  \includegraphics*[width=0.44\textwidth,clip]{03.eps}\hspace{5mm}
  \includegraphics*[width=0.44\textwidth,clip]{04.eps}\\\vspace{2mm}
  \includegraphics*[width=0.44\textwidth,clip]{05.eps}\hspace{5mm}
  \includegraphics*[width=0.44\textwidth,clip]{06.eps}\\\vspace{2mm}
  \includegraphics*[width=0.44\textwidth,clip]{07.eps}
  \caption{Potential energy curves (PECs) of hydrogen-bonded
    duplexes. The S22 data points (cross marks) are taken from
    T.~Takatani, E.~G.~Hohenstein, M. Malagoli, M.~S.~Marshall, and
    C.~D.~Sherrill, J. Chem. Phys. {\bf 132}, 144104 (2010). The
    CCSD(T) PECs data (dashed lines with open circles) in this and
    subsequent figures are taken from L.~F. Molnar, X. He, B. Wang,
    and J. Kenneth M.~Merz, J. Chem.\ Phys.~{\bf 131}, 065102
    (2009). For the hydrogen-bonded adenine-thymine duplex, CCSD(T)
    PEC data is not available. The shapes near minima are compared in
    inset figures where PECs are aligned to have the common minimum
    point.}
  \label{fig:h-pecs}
\end{figure*}

\begin{figure*}[!ht]
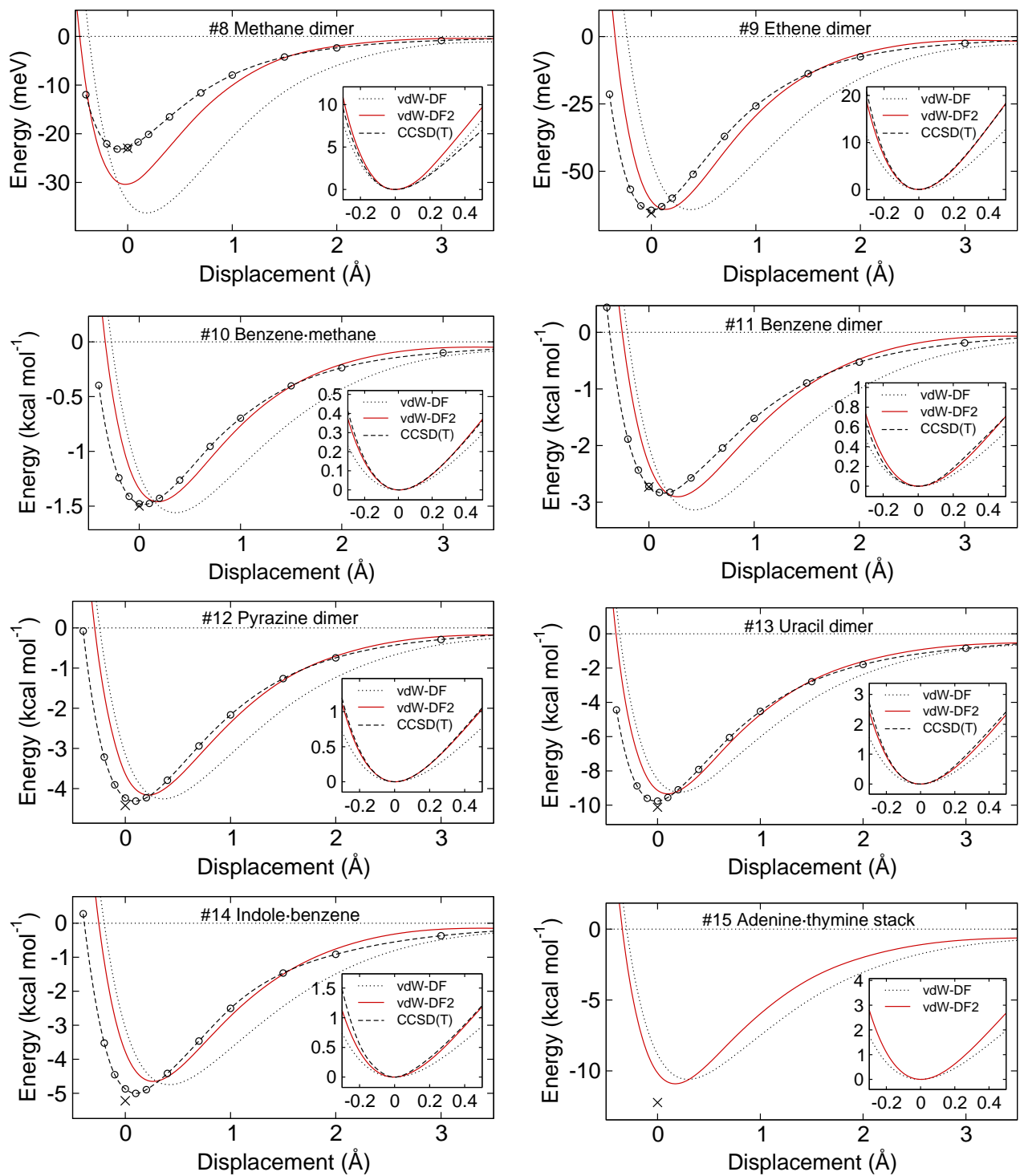

  \centering
  \includegraphics*[width=0.45\textwidth,clip]{08.eps}\hspace{5mm}
  \includegraphics*[width=0.45\textwidth,clip]{09.eps}\\\vspace{2mm}
  \includegraphics*[width=0.45\textwidth,clip]{10.eps}\hspace{5mm}
  \includegraphics*[width=0.45\textwidth,clip]{11.eps}\\\vspace{2mm}
  \includegraphics*[width=0.45\textwidth,clip]{12.eps}\hspace{5mm}
  \includegraphics*[width=0.45\textwidth,clip]{13.eps}\\\vspace{2mm}
  \includegraphics*[width=0.45\textwidth,clip]{14.eps}\hspace{5mm}
  \includegraphics*[width=0.45\textwidth,clip]{15.eps}
  \caption{Potential energy curves (PECs) of dispersion-dominated
    duplexes.  For the stacked adenine-thymine duplex, CCSD(T) PEC is
    not available.  The S22 data points (cross marks) are taken from
    T.~Takatani, E.~G.~Hohenstein, M. Malagoli, M.~S.~Marshall, and
    C.~D.~Sherrill, J. Chem. Phys. {\bf 132}, 144104 (2010).  The
    original S22 values (plus marks taken from P.~Jurecka, J.~Sponer,
    J.~Cern\'{y}, and P.~Hobza, Phys. Chem. Chem. Phys. {\bf 8}, 1985
    (2006)) are also shown if the difference between those two values
    are larger than 3\%.}
  \label{fig:vdw-pecs}
\end{figure*}

\begin{figure*}[!ht]
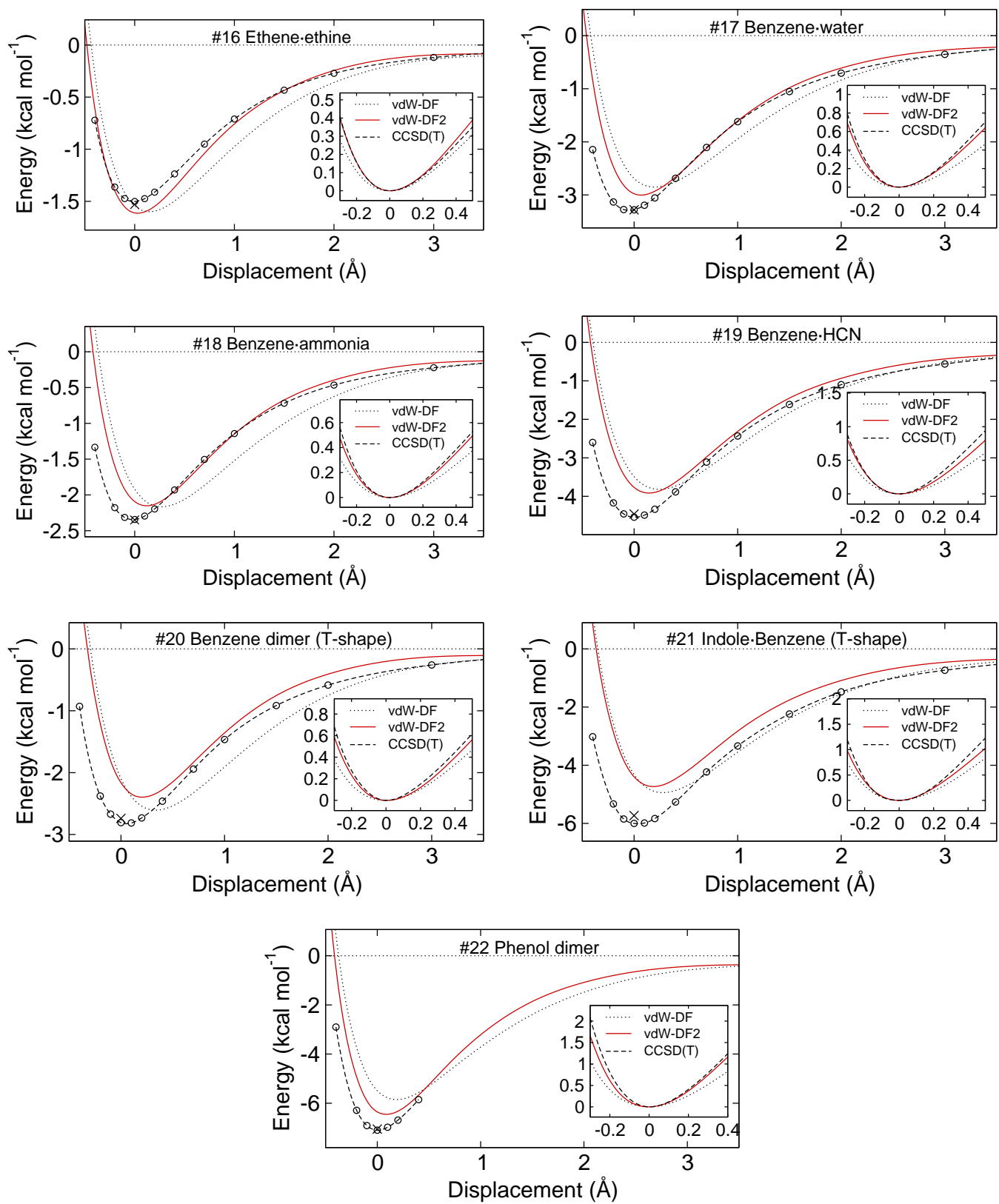

  \centering
  \includegraphics*[width=0.48\textwidth,clip]{16.eps}\hspace{5mm}
  \includegraphics*[width=0.48\textwidth,clip]{17.eps}\\\vspace{5mm}
  \includegraphics*[width=0.48\textwidth,clip]{18.eps}\hspace{5mm}
  \includegraphics*[width=0.48\textwidth,clip]{19.eps}\\\vspace{5mm}
  \includegraphics*[width=0.48\textwidth,clip]{20.eps}\hspace{5mm}
  \includegraphics*[width=0.48\textwidth,clip]{21.eps}\\\vspace{5mm}
  \includegraphics*[width=0.48\textwidth,clip]{22.eps}
  \caption{Potential energy curves (PECs) of mixed interaction
    duplexes.}
  \label{fig:mixed-pecs}
\end{figure*}

\begin{figure*}[!ht]
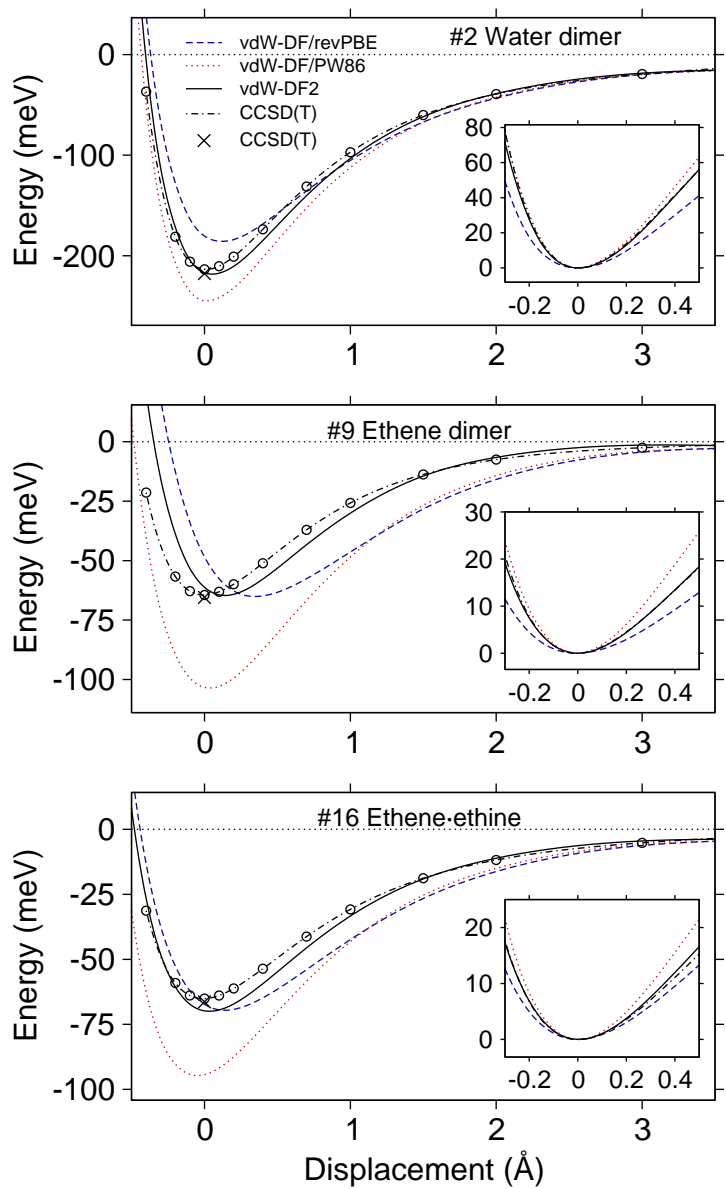

  \centering
  \includegraphics*[width=0.53\textwidth,clip]{comp02.eps}\\\vspace{3mm}
  \includegraphics*[width=0.53\textwidth,clip]{comp09.eps}\\\vspace{3mm}
  \includegraphics*[width=0.53\textwidth,clip]{comp16.eps}
  \caption{Comparison of vdW-DF with revPBE exchange functional
    (vdW-DF/revPBE), vdW-DF with PW86R (vdW-DF/PW86R), and vdW-DF2
    with PW86R (vdW-DF2).  Potential energy curves (PECs) of water
    dimer (hydrogen-bonded duplexes), ethene dimer (dispersion),
    ethene-ethine duplex (mixed) are shown. Note that the difference
    between vdW-DF/revPBE and vdW-DF/PW86R vanishes beyond 1
    \AA{}. Replacing revPBE exchange with PW86 weakens the repulsion
    near equilibrium separations and gives more accurate separations
    but also large overbinding. Furthermore, the comparison of PECs
    with CCSD(T) shows that the original vdW-DF consistently
    overestimates the vdW attraction. The use of large-N asymptote in
    determining the internal functional for vdW kernel effectively
    weakens the vdW attraction and vdW-DF2 gives an excellent
    agreement with CCSD(T) for all separations.}
  \label{fig:comp-pw86}
\end{figure*}

\begin{table*}[!ht]
  \centering
  \caption{Graphite interlayer binding energy and interlayer spacing
    (i.e., half of the $c$ lattice constant). An AB stacked 4-layer
    unit cell is used for vdW-DF and vdW-DF2 calculations. The
    interlayer binding energy is calculated by subtracting the energy
    of each single layer in the same unit cell from that of 4-layer
    infinite bulk.}
  \label{tab:graphite}
  \begin{ruledtabular}
    \begin{tabular}{lcccc}
       & 
      \multicolumn{1}{c}{vdW-DF} & 
      \multicolumn{1}{c}{vdW-DF2} &
      \multicolumn{1}{c}{QMC\footnotemark[1]} &
      \multicolumn{1}{c}{Experiment\footnotemark[2]} \\ \hline
      Binding energy (meV/atom) &  50  &  49  & 56$\pm$5 & 52$\pm$5 \\
      Interlayer spacing (\AA)  & 3.60 & 3.53 &   3.43   & 3.36 \\
    \end{tabular}
  \end{ruledtabular}
  \footnotetext[1]{L. Spanu, S. Sorella, and G. Galli,
    Phys. Rev. Lett. \textbf{103}, 196401 (2009). Zero-point energy and
    phonon contributions at 300K are included.}
  \footnotetext[2]{R. Zacharia, H. Ulbricht, and T. Hertel,
    Phys. Rev. B \textbf{69}, 155406 (2004).}
\end{table*}

\begin{figure*}[tbp!]
  \centering
  \includegraphics[width=3in,clip=true]{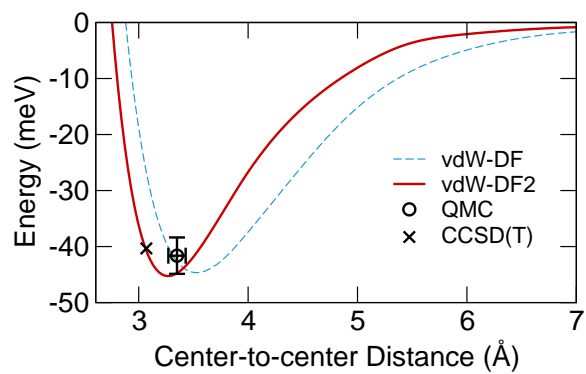}
  \caption{Hydrogen-benzene potential energy curve (PEC). Sampled
    along a normal vector of the benzene plane passing through the
    benzene center. H$_2$ is parallel to the vector. The CCSD(T)
    result is taken from H\"{u}bner \textit{et al.}  J. Phys. Chem. A
    \textbf{108}, 3019 (2004) and QMC from Beaudet \textit{et al.}
    J. Chem. Phys. \textbf{129}, 164711 (2008).}
  \label{fig:h2bz} 
\end{figure*}

\begin{figure*}[!ht]
  \centering
  \includegraphics[width=\textwidth,clip=true]{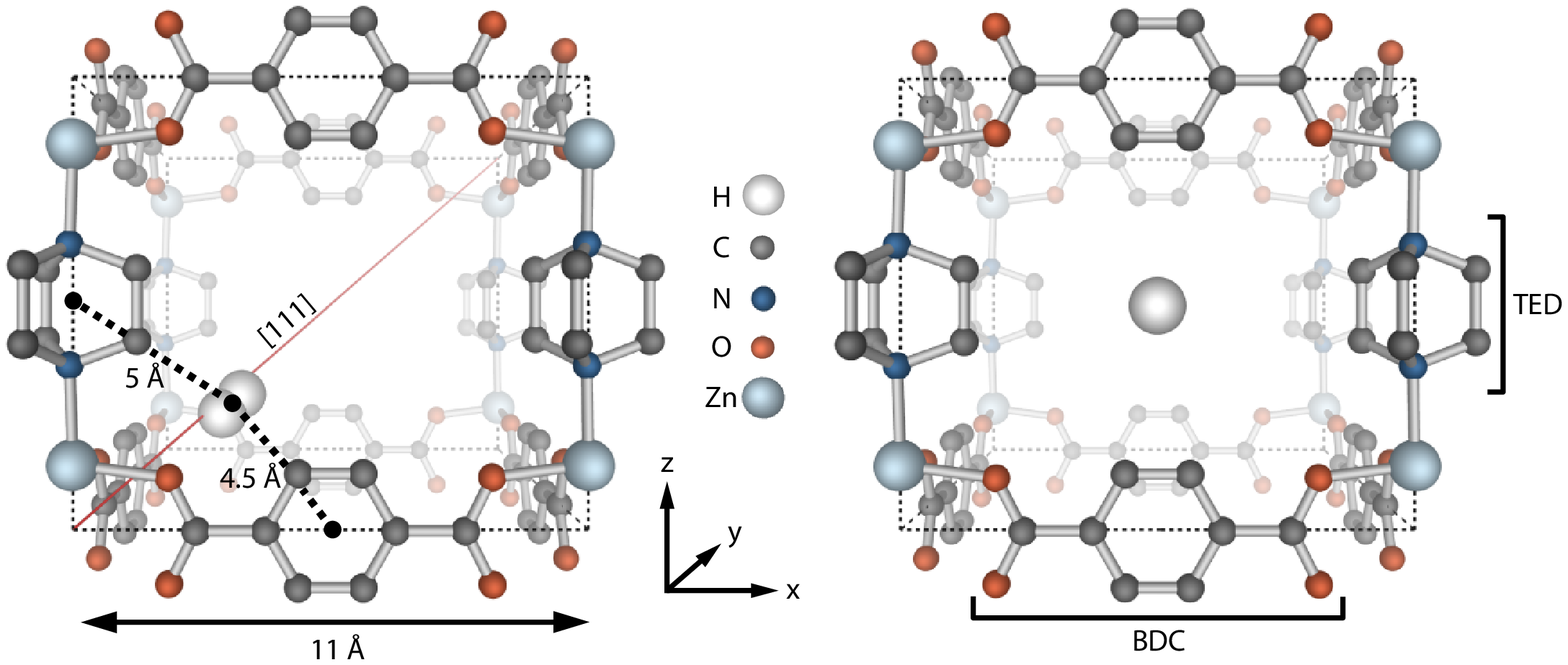}
  \caption{Hydrogen adsorption by the metal-organic framework (MOF)
    structure Zn$_2$(BDC)$_2$(TED) where BDC and TED denote the linking 
    structures so labeled above. Hydrogen atoms attached to carbon
    atoms are removed for simplicity. The left panel shows the H$_2$
    in a ``corner'' site along the body diagonal, and the right panel
    shows the H$_2$ head-on in the ``face center''  site. The binding
    of the H$_2$ is dominated in both cases by its interactions
    with the benzene and TED complexes which are $\sim$5 {\AA} away.} 
  \label{fig:ted} 
\end{figure*}

\begin{table}[!ht]
  \centering
  \caption{Hydrogen binding energies in Zn$_2$(BDC)$_2$(TED) (in meV).
    Zero-point energies are included in the calculated values. These
    values demonstrate the importance for the functional to give
    accurate interaction energies at intermediate distances that are
    longer than the equilibrium distances in isolated duplexes. For
    example, the hydrogen-benzene interaction (Fig.~\ref{fig:h2bz}) at
    a center-to-center distance of 5 \AA{} is $\sim$100\%
    overestimated by vdW-DF, although the binding energy at the
    equilibrium separation is quite accurate. This effect leads to the
    60\% overestimate of the binding of the most strongly bound H$_2$
    in the Zn$_2$(BDC)$_2$(TED) crystal by vdW-DF. In vdW-DF2 the
    error reduced to a reasonable size, and a switching of which site
    is the most strongly bound also occurs.}
  \label{tab:ted}
  \begin{ruledtabular}
    \begin{tabular}{cccc}
      \multirow{2}{*}{H$_2$ binding site} & \multicolumn{3}{c}{H$_2$
        binding energy (meV)} \\
      \cline{2-4}
      & vdW-DF\footnotemark[1] & vdW-DF2 & Expt.\footnotemark[2] \\
      \hline
      Corner site      & 74 & 65 & \multirow{2}{*}{52} \\
      Face-center site & 82 & 56 & \\
    \end{tabular}
  \end{ruledtabular}
  \footnotetext[1]{
    L. Kong, V. R. Cooper, N. Nijem, K. Li, J. Li, Y. J. Chabal, and D.
    C. Langreth, Phys. Rev. B \textbf{79}, 081407(R) (2009).
  }
  \footnotetext[2]{
    J. Y. Lee, D. H. Olson, L. Pan, T. J. Emge, and J. Li,
    Adv. Funct. Mater. \textbf{17}, 1255 (2007).
  }
\end{table}

\begin{figure*}[!ht]
  \centering
  \includegraphics[width=6in,clip=true]{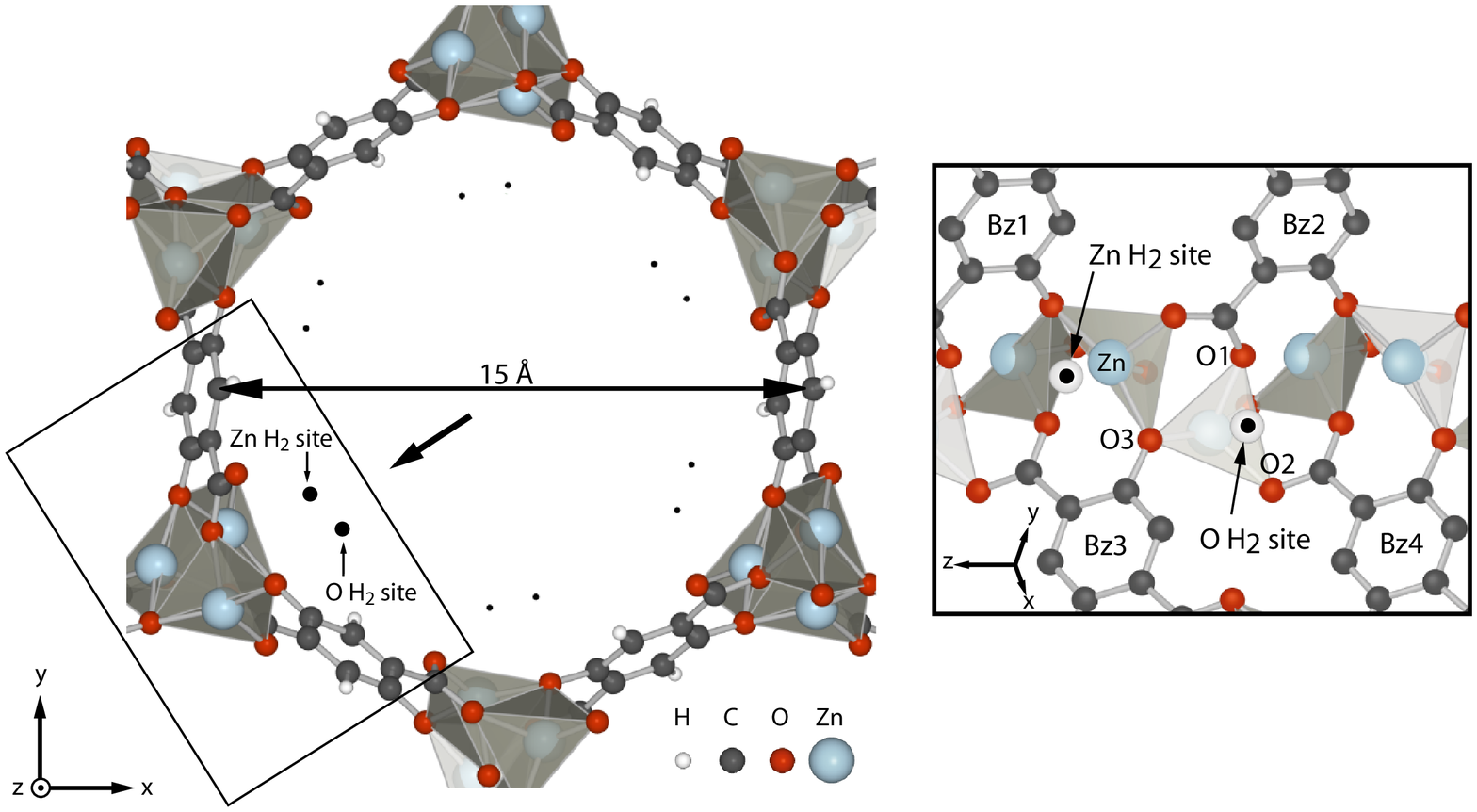}
  \caption{Hydrogen adsorption sites (black dots) in MOF-74
    [N. L. Rosi, J. Kim, M. Eddaoudi, B. Chen, M. O'Keeffe, and
    O. M. Yaghi, J. Am. Chem. Soc. \textbf{127}, 1504 (2005)]. (Left)
    Top view into the pore. (Right) Side view of the pore wall. The
    viewing direction and area are depicted by a short arrow and a
    rectangle, respectively, in the left figure.}
  \label{fig:mof74} 
\end{figure*}

\begin{table*}[!ht]
  \centering
  \caption{Equilibrium separations (in \AA) of hydrogens to the
    nearest atom, to the nearest hydrogen, and to nearby benzene (Bz)
    rings in MOF-74. Calculated binding energies (including zero-point
    energies) of H$_2$ at Zn site are given at the last line. In this
    MOF, the binding is dominated by the proximity of the H$_2$ to the
    unsaturated Zn atom, and the accurate binding energy predicted by
    vdW-DF is essentially unchanged in vdW-DF2. Still, the weakening
    of the intermediate range attraction is presumably compensated by
    the closer proximity of the H$_2$ to the Zn atom in vdW-DF2.}
  \label{tab:mof74}
  \begin{ruledtabular}
    \begin{tabular}{ccccc}
      \multirow{2}{*}{From} &
      \multirow{2}{*}{To\footnotemark[3]} & 
      \multicolumn{3}{c}{Distance (\AA)} \\
      \cline{3-5}
      & & vdW-DF\footnotemark[1] & vdW-DF2 & Expt.\footnotemark[2]\\
      \hline
      \multirow{4}{*}{Zn H$_2$ site}
      & Zn            & 3.0 & 2.8 & 2.6 \\
      & Bz1\footnotemark[4]           & 4.2 & 4.1 & 3.9 \\
      & Bz3\footnotemark[4]           & 4.4 & 4.3 & 4.2 \\
      &  O H$_2$ site & 3.3 & 3.1 & 2.9 \\
      \hline
      \multirow{4}{*}{O H$_2$ site}
      &  O1 & 3.7 & 3.4 & 3.3 \\
      &  O2 & 4.3 & 4.0 & 3.5 \\
      &  O3 & 3.4 & 3.3 & 3.1 \\
      & Bz2\footnotemark[4] & 5.1 & 4.8 & 5.0 \\
      & Bz3\footnotemark[4] & 4.7 & 4.7 & 4.6 \\
      & Bz4\footnotemark[4] & 5.4 & 5.3 & 4.7 \\
      \hline
      \multicolumn{2}{c}{H$_2$ binding energy at Zn site (eV)}
      & 0.10           & 0.10           &  0.09   \\
    \end{tabular}
  \end{ruledtabular}
  \footnotetext[1]{
    L. Kong, G. Román-Pérez, J. M. Soler, and D. C. Langreth,
    Phys. Rev. Lett. \textbf{103}, 096103 (2009).
  }
  \footnotetext[2]{
    Y. Liu, H. Kabbour, C. M. Brown, D. A. Neumann, and C. C. Ahn,
    Langmuir \textbf{24}, 4772 (2008).
  }
  \footnotetext[3]{
    See the right panel of Fig.~\ref{fig:mof74} for a depiction of the
    corresponding atomic locations.
  }
  \footnotetext[4]{
    To the center of the molecule.
  }
\end{table*}

\begin{figure*}[!ht]
  \centering
  \includegraphics[height=2.5in,clip=true]{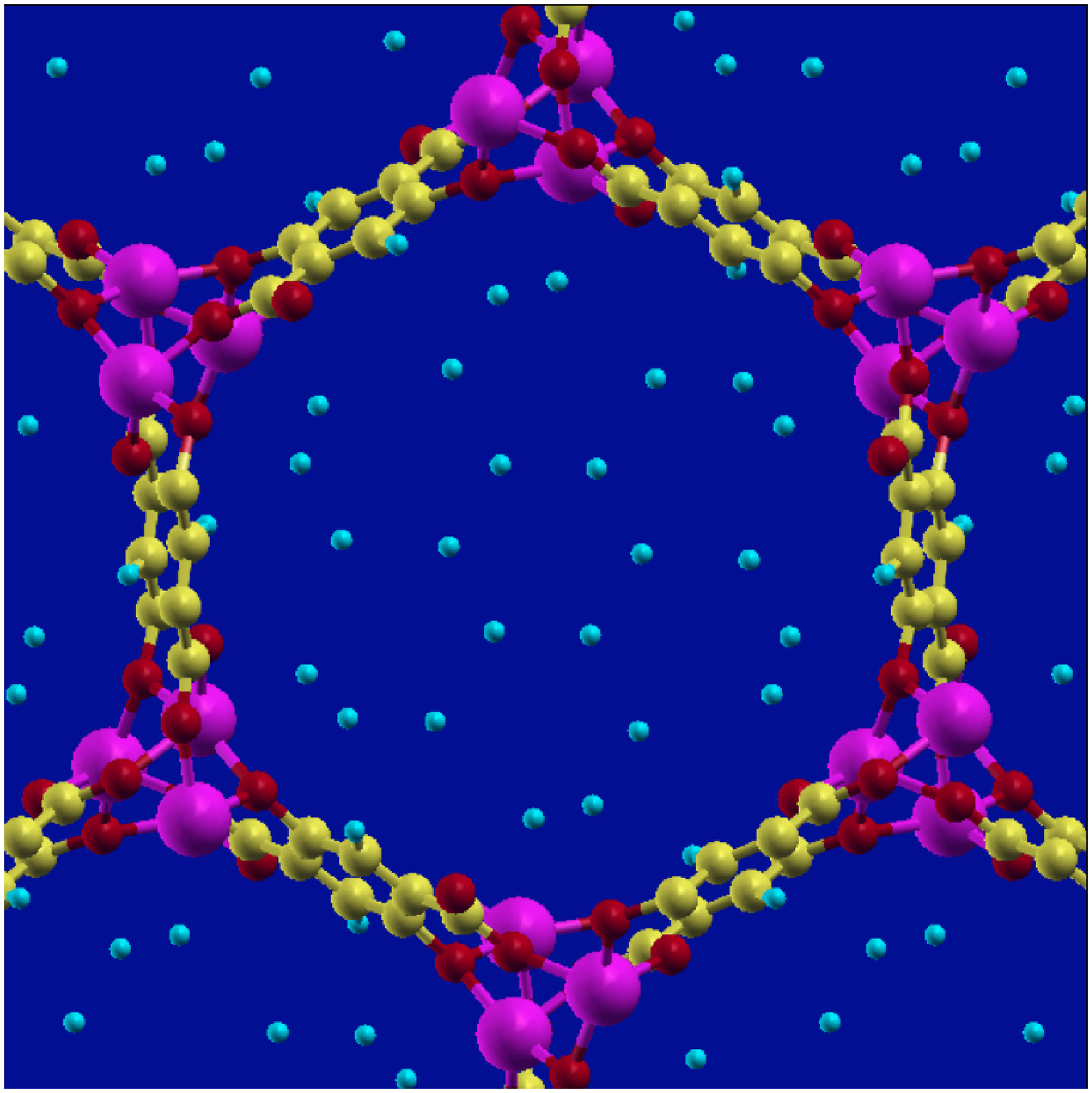}
  \includegraphics[height=2.5in,clip=true]{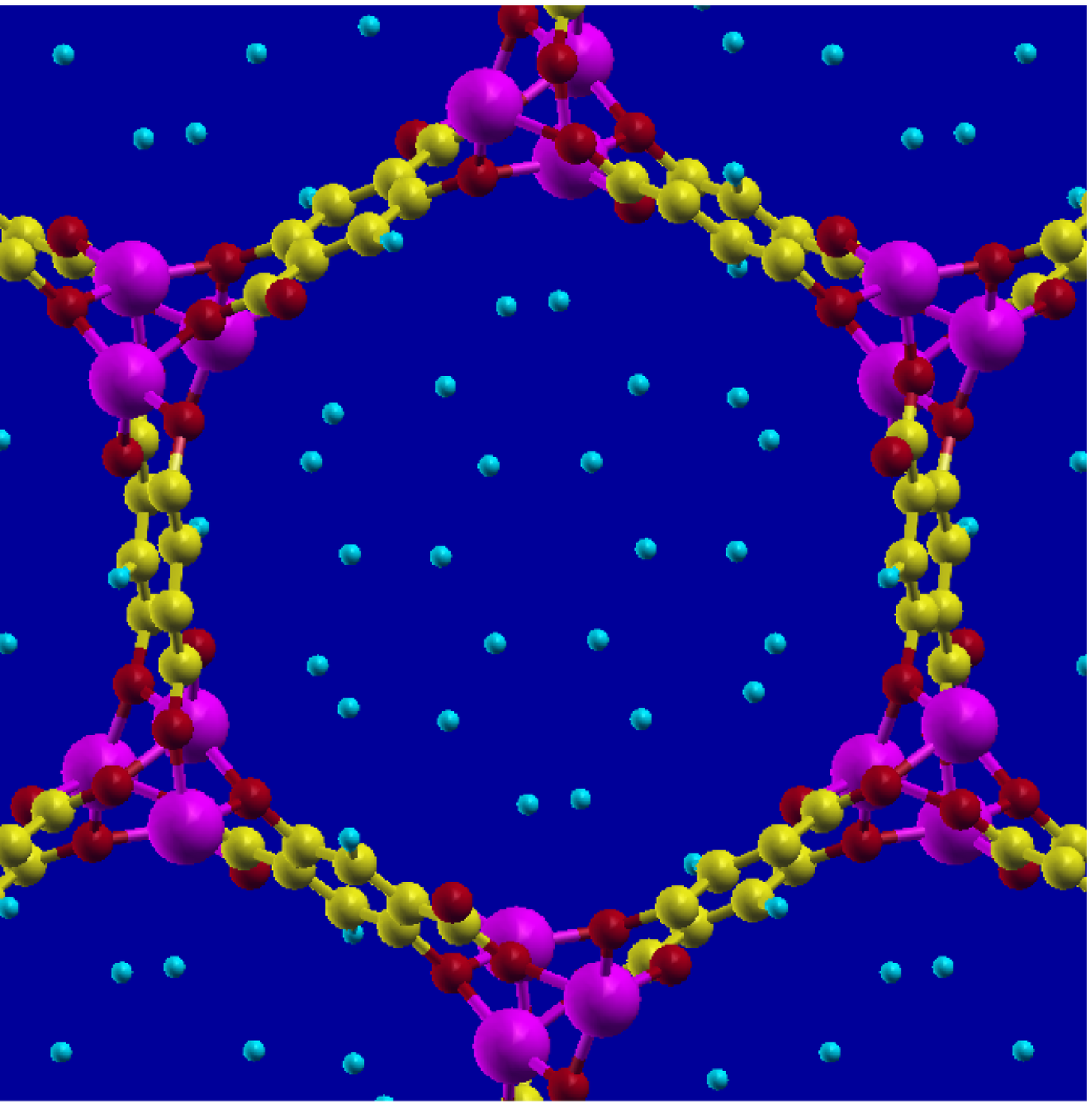}
  \caption{D$_2$ adsorption sites in the fully loaded MOF-74: Via
    neutron diffraction$^a$ (left panel); Via vdW-DF2 (right panel).
  }
  $^a${Y. Liu, H. Kabbour, C. M. Brown, D. A. Neumann, and C. C. Ahn,
    Langmuir \textbf{24}, 4772 (2008).}
  \label{fig:loaded} 
\end{figure*}

\end{document}